\documentclass[10pt,conference]{IEEEtran}
\usepackage{subfigure}
\usepackage{graphicx}
\usepackage{amsmath}
\usepackage{epsfig}
\usepackage{amsfonts}
\usepackage{amssymb}
\usepackage{gensymb}
\usepackage{epstopdf}

\newcommand{\argmin}{\mathop{\text{argmin}}}

\newcommand{\vx}{{\bf x}}
\newcommand{\vy}{{\bf y}}

\newcommand{\vn}{{\bf n}}

\newcommand{\mh}{{\bf H}}

\newcommand{\gsm}{{\mathbb S}_{N_t,{\mathbb M}}^{N_{a}}}
\newcommand{\sm}{{\mathbb M}}
\newcommand{\se}{{\mathbb E}}

\newcommand{\Define}{\triangleq}

\begin{document}
\title{{\huge 
Generalized Spatial Modulation in Indoor Wireless Visible Light 
Communication}}
\author{
S. P. Alaka, T. Lakshmi Narasimhan, and A. Chockalingam \\
Department of Electrical Communication Engineering \\
Indian Institute of Science, Bangalore, India
}

\maketitle

\begin{abstract}
In this paper, we investigate the performance of generalized spatial 
modulation (GSM) in indoor wireless visible light communication (VLC) 
systems. GSM uses $N_t$ light emitting diodes (LED), but activates only 
$N_a$ of them at a given time. Spatial modulation and spatial multiplexing 
are special cases of GSM with $N_{a}=1$ and $N_{a}=N_t$, respectively. 
We first derive an analytical upper bound on the bit error rate (BER) for 
maximum likelihood (ML) detection of GSM in VLC systems. Analysis and
simulation results show that the derived upper bound is very tight at 
medium to high signal-to-noise ratios (SNR). The channel gains and channel 
correlations influence the GSM performance such that the best BER is 
achieved at an optimum LED spacing. Also, for a fixed transmission 
efficiency, the performance of GSM in VLC improves as the half-power 
semi-angle of the LEDs is decreased. We then compare the performance 
of GSM in VLC systems with those of other MIMO schemes such as spatial 
multiplexing (SMP), space shift keying (SSK), generalized space shift 
keying (GSSK), and spatial modulation (SM). Analysis and simulation 
results show that GSM in VLC outperforms the other considered MIMO 
schemes at moderate to high SNRs; for example, for 8 bits per channel 
use, GSM outperforms SMP and GSSK by about 21 dB, and SM by about 
10 dB at $10^{-4}$ BER. 
\end{abstract}


\section{Introduction}
\label{sec1}
The radio frequency (RF) spectrum used in industrial, scientific and medical
radio bands and telecommunication radio bands are crowded with various 
wireless communication systems. Recently, optical wireless communication 
technology, where information is conveyed through optical radiations in
free space in outdoor and indoor environments, is emerging as a promising 
complementary technology to RF communication technology. While communication 
using infrared wavelengths has been in existence for quite some time
\cite{chann1},\cite{chann2}, more recent interest centers around 
indoor communication using visible light wavelengths 
\cite{haas1},\cite{brien}. A major attraction in indoor visible light 
communication (VLC) is the potential to simultaneously provide both
energy-efficient lighting as well as high-speed short-range communication 
using inexpensive high-luminance light-emitting diodes (LED). Several 
other advantages including no RF radiation hazard, abundant VLC spectrum 
at no cost, and very high data rates make VLC increasingly popular. For 
example, a 3 Gbps single-LED VLC link based on OFDM has been reported 
recently \cite{haas2}. Also, multiple-input multiple-output (MIMO) 
techniques, which are immensely successful and popular in RF 
communications \cite{mimo1},\cite{mimo2}, can be employed in VLC
systems to achieve improved communication efficiencies 
\cite{jean},\cite{vlc5},\cite{vlc6}. In particular, it has been 
shown that MIMO techniques can provide gains in VLC systems even
under line-of-sight (LOS) conditions which provide only little 
channel differences \cite{vlc5}. Our new contribution in this 
paper is the investigation of {\em generalized spatial modulation 
(GSM)}, an attractive MIMO transmission scheme, in the context of
VLC. Such a study, to our knowledge, has not been reported before. 

In the context of VLC systems, MIMO techniques including spatial 
multiplexing (SMP), space shift keying (SSK), generalized space shift 
keying (GSSK), and spatial modulation (SM) have been investigated in
the literature \cite{vlc5}-\cite{vlc2}. In SMP, there are $N_t$ LEDs 
at the transmitter and all of them are activated simultaneously in a 
given channel use, such that $N_t$ symbols from a positive real-valued
$|\sm|$-ary pulse amplitude modulation (PAM) alphabet $\sm$ are sent in 
a channel use \cite{vlc5}. Thus, the transmission efficiency in SMP is 
$\eta_{smp}=N_t\lfloor \log_2|\sm|\rfloor$ bits per channel use 
(bpcu). In SSK, there are $N_t$ LEDs, out of which only one will be 
activated in a given channel use \cite{vlc4}. The LED to be activated 
is chosen based on $\lfloor\log_2N_t\rfloor$ information bits. Only the 
index of this active LED will convey information bits, so that the 
transmission efficiency is $\eta_{ssk}=\lfloor\log_2N_t\rfloor$ bpcu. 
This means that a large number of LEDs is needed to achieve high 
transmission efficiencies in SSK. That is, since 
$N_t=\lceil2^{\eta_{ssk}}\rceil$, the number of LEDs required in SSK 
is exponential in the transmission efficiency $\eta_{ssk}$. On the other 
hand, SSK has the advantage of having no interference, since only one 
LED will be active at any given time and the remaining LEDs will be OFF.
GSSK is a generalization of SSK, in which $N_a$ out of $N_t$ LEDs will 
be activated in a given channel use, and the indices of the active LEDs 
will convey information bits \cite{vlc3}-\cite{vlc3b}. Since there are 
$N_t \choose N_a$ possibilities of choosing the active LEDs, the 
transmission efficiency in GSSK is given by 
$\eta_{gssk}=\lfloor\log_2{N_t \choose N_a}\rfloor$ bpcu. 

SM is similar to SSK (i.e., one out of $N_t$ LEDs is activated and this
active LED is chosen based on $\lfloor\log_2 N_t\rfloor$ information bits),
except that in SM a symbol from a positive real-valued $|\sm|$-ary PAM 
alphabet $\sm$ is sent on the active LED. So, the transmission efficiency 
in SM is $\eta=\lfloor\log_2 N_t\rfloor+\lfloor\log_2 |\sm|\rfloor$
bpcu. A comparative study of SMP and SM in VLC systems has shown that, for 
the same transmission efficiency, SM outperforms SMP under certain geometric 
conditions \cite{vlc5}.  

Like the generalization of SSK to GSSK, it is possible to generalize SM. 
That is, activate $N_a$ out of $N_t$ LEDs in a given channel use, and, 
on each active LED, send a symbol from a positive real-valued $|\sm|$-ary
PAM alphabet $\sm$. Such a scheme, referred to as {\em generalized spatial 
modulation (GSM)}, then has a transmission efficiency of 
$\eta_{gsm}=\lfloor\log_2{N_t \choose N_a}\rfloor+N_a\lfloor\log_2 |\sm|\rfloor$ 
bpcu. Note that both SM and SMP become special cases of GSM for $N_a=1$ 
and $N_a=N_t$, respectively. GSM in the context of RF communications has 
been investigated in the literature \cite{gsm1}-\cite{gsm3}. However, GSM 
in the context of VLC systems has not been reported so far. Our contribution 
in this paper attempts to fill this gap. In particular, we investigate, 
through analysis and simulations, the performance of GSM in comparison with 
other MIMO schemes including SMP, SSK, GSSK, and SM. Our performance study 
reveals favorable results for GSM compared to other MIMO schemes.

The rest of this paper is organized as follows. In Sec. \ref{sec2}, we 
present the considered indoor VLC system model. In Sec. \ref{sec3}, we 
present the GSM scheme for VLC. In Sec. \ref{sec4}, we derive an upper 
bound on the bit error probability of GSM for maximum likelihood
(ML) detection in VLC. In Sec. \ref{sec5}, we present a detailed 
performance comparison between GSM and other MIMO schemes in VLC. 
Finally, conclusions are presented in Sec. \ref{sec6}.

\section{System model}
\label{sec2}
\begin{figure}
\centering
\includegraphics[height=1.2in]{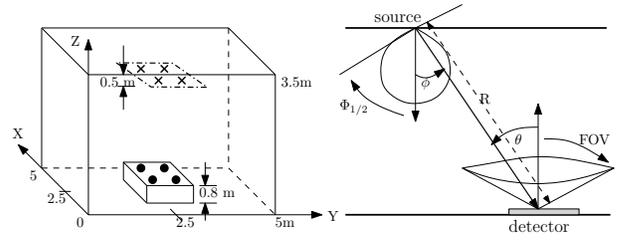}
\caption{Geometric set-up of the considered indoor VLC system. 
A dot represents a photo detector and a cross represents an LED.}
\label{sys}
\vspace{-2mm}
\end{figure}

Consider an indoor VLC system with $N_t$ LEDs (transmitter) and $N_r$ 
photo detectors (receiver). We assume that the LEDs have a Lambertian 
radiation pattern \cite{chann2},\cite{new1}. In a given channel use, 
each LED is either OFF or emits light of some positive intensity 
$I \in \mathbb M$, where $\mathbb M$ is the set of all possible 
intensity levels. An LED which is OFF is considered to send a signal of 
intensity zero. Let $\vx$ denote the $N_t\times 1$ transmit signal vector, 
where the $i$th element of $\bf x$ is $x_i\in\{{\mathbb M} \cup 0\}$. Let 
$\mh$ denote the $N_r\times N_t$ optical MIMO channel matrix, given by
\begin{eqnarray}
\mh=
\begin{bmatrix}
h_{11} & h_{12} & h_{13} & \cdots & h_{1N_t} \\
h_{21} & h_{22} & h_{23} & \cdots & h_{2N_t} \\
\vdots & \vdots & \ddots & \vdots & \vdots \\
h_{N_r1}& h_{N_r2} & h_{N_r3} & \cdots & h_{N_rN_t}
\end{bmatrix},
\end{eqnarray}
where $h_{ij}$ is the channel gain between $j$th LED and $i$th photo detector,
$j=1,2,\cdots,N_t$ and $i=1,2,\cdots,N_r$. As in \cite{vlc5}, we consider 
only the line-of-sight (LOS) paths between the LEDs and the photo detectors,
and assume no time-dispersion (because of negligible path delay differences
between LEDs and photo detectors). From \cite{chann2}, the LOS channel gain 
$h_{ij}$ is calculated as (see Fig. \ref{sys} for the definition of various 
angles in the model)
\begin{equation}
{h_{ij}} = \frac{n+1}{2\pi}\cos^{n}{\phi_{ij}}\, 
\cos{\theta_{ij}}\frac{A}{R_{ij}^2} 
\mbox{rect}\Big(\frac{\theta_{ij}}{FOV}\Big),
\label{channel}
\end{equation}
where $\phi_{ij}$ is the angle of emergence with respect to the $j$th
source (LED)
and the normal at the source, $n$ is the mode number of the radiating lobe
given by 
\[
n=\frac{-\ln(2)}{\ln\cos{\Phi_{\frac{1}{2}}}},
\] 
$\Phi_\frac{1}{2}$ is the half-power semiangle of the LED \cite{new1}, 
$\theta_{ij}$ is the angle of incidence at the $i$th photo detector, 
$A$ is the area of the detector, $R_{ij}$ is the distance between the 
$j$th source and the $i$th detector, FOV is the field of view of the 
detector, and 
\[
\mbox{rect}(x)=
\begin{cases}
1, & |x|\leq 1 \\
0, & |x|>1.
\end{cases}\]
	
\begin{figure}[h]
\centering
\subfigure[Transmitter, $N_t=4$]{
\includegraphics[height=1in]{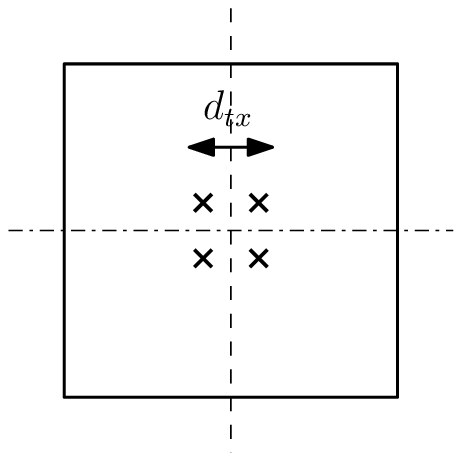}
\label{placement1}
}
\subfigure[Receiver, $N_r=4$]{
\includegraphics[height=1in]{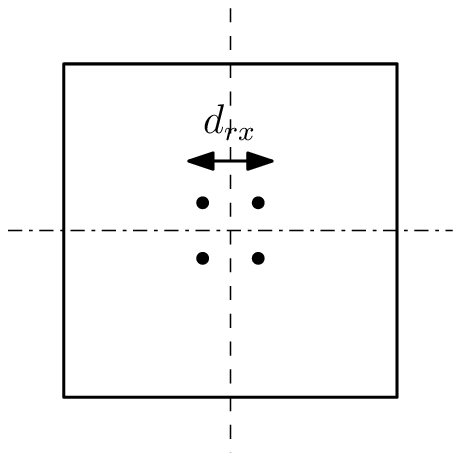}
\label{placement2}
}
\subfigure[\hspace{-1mm}Transmitter, $N_t$=16]{
\includegraphics[height=1in]{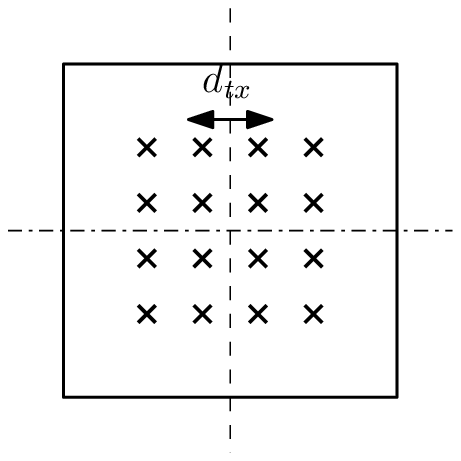}
\label{placement3}
}
\caption{Placement of LEDs and photo detectors.}
\label{place}
\end{figure}

The LEDs and the photo detectors are placed in a room of size
5m$\times$5m$\times$3.5m as shown in Fig. \ref{sys}. The LEDs are 
placed at a height of 0.5m below the ceiling and the photo detectors 
are placed on a table of height 0.8m. Let $d_{tx}$ denote the distance 
between the LEDs and $d_{rx}$ denote the distance between the photo 
detectors (see Fig. \ref{place}). We choose $d_{tx}$ as 0.6m and $d_{rx}$ 
as 0.1m. For example, when $N_t=N_r=4$, the placement of LEDs and photo
detectors is depicted in Figs. \ref{placement1},\ref{placement2}. 
When $N_t=16$, the placement of LEDs is depicted in Fig. \ref{placement3}.

Assuming perfect synchronization, the $N_r\times 1$ received signal vector 
at the receiver is given by
\begin{eqnarray}
\vy = r\mh\vx+\vn,
\label{sysmodel}
\end{eqnarray}
where 
$\vx$ is an $N_t$-dimensional vector with exactly $N_a$ non-zero elements such
that each element in $\vx$ belongs to $\{{\mathbb M} \cup 0\}$,
$r$ is the responsivity of the detector \cite{new2} and $\vn$ is 
the noise vector of dimension $N_r\times 1$. Each element in the noise 
vector $\vn$ is the sum of received thermal noise and ambient shot light 
noise, which can be modeled as i.i.d. real AWGN with zero mean and variance 
$\sigma^2$ \cite{chann1}. The average received signal-to-noise ratio (SNR) 
is given by 
\begin{eqnarray}
{\overline{\gamma}}&=&\frac{r^2P_r^2}{\sigma^2},
\end{eqnarray}
where 
$P_r^2 = \frac{1}{N_r}\sum\limits_{i=1}^{N_r}\se [{|\mh_i\vx|}^2]$, 
and $\mh_i$ is the $i$th row of $\mh$. 

\section{GSM in VLC systems}
\label{sec3}
In GSM, information bits are conveyed not only through modulation symbols
sent on active LEDs, but also through indices of the active LEDs. In each 
channel use, the transmitter selects $N_a$ out of $N_t$ LEDs to activate. 
This selection is done based on $\lfloor\log_2{N_t \choose N_{a}}\rfloor$ 
information bits. Each active LED emits an $M$-ary intensity modulation 
symbol $I \in \sm$, where $\sm$ is the set of intensity levels given by
\cite{vlc5}
\begin{equation}
{I_m}=\frac{2I_pm}{M+1}, \quad m = 1,2,\cdots,M, 
\label{intensities}
\end{equation} 
where $M\Define |\sm|$ and $I_p$ is the mean optical power emitted. 
Therefore, the total number 
of bits conveyed in a channel use in GSM is given by
\begin{equation}
\eta_{gsm}=\left\lfloor\log_2{N_t \choose N_{a}}\right\rfloor+N_{a}\left\lfloor\log_2M\right\rfloor \quad \mbox{bpcu}.
\label{eta}
\end{equation}

Let $\gsm$ denote the GSM signal set, which is the set of all possible 
GSM signal vectors that can be transmitted. Out of the $N_t\choose N_{a}$ 
possible LED activation patterns\footnote{LED activation pattern is a 
$N_a$-tuple of the indices of the active LEDs in any given channel use.}, 
only $2^{\lfloor \log_2{N_t\choose N_{a}}\rfloor}$ activation patterns are 
needed for signaling. 

\begin{figure}
\centering
\includegraphics[height=1.9in]{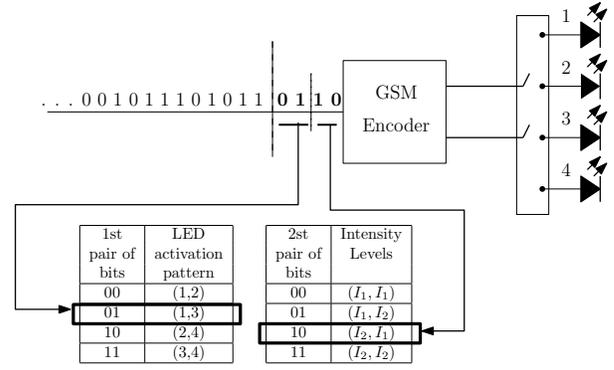}
\caption{GSM transmitter for VLC system with $N_t=4, N_a=2, M=2$.}
\label{GSM}
\vspace{-2mm}
\end{figure}

{\em Example 1:} Let $N_t=4$ and $N_a=2$. In this configuration, the number 
of bits that can be conveyed through the LED activation pattern is 
$\lfloor\log_2{4 \choose 2}\rfloor$ = 2 bits. Let the number of intensity 
levels be $M = 2$, where $I_1=\frac{2}{3}$ and $I_2=\frac{4}{3}$. 
This means that one bit on each of the active LED is sent through intensity 
modulation. Therefore, the overall transmission efficiency is 4 bpcu. 
In each channel use, four bits from the incoming bit stream are transmitted.
Of the four transmitted bits, the first two correspond to the LED activation 
pattern and the next two bits correspond to the intensity levels of the 
active LEDs. This GSM scheme is illustrated in Fig. \ref{GSM}, where the
first two bits `01' choose the active LEDs pair $(1,3)$ and the second two 
bits `10' choose the intensity levels $(I_2,I_1)$, where LED 1 emits 
intensity $I_2$, LED 3 emits intensity $I_1$, and the other LEDs remain 
inactive (OFF). In this example, we require only 4 activation patterns out 
of ${4\choose 2}=6$ possible activation patterns. So the GSM signal set for 
this example can be chosen as follows: 

{\small
\begin{eqnarray}
\hspace{-0mm}
{\mathbb S}_{4,2}^2
\hspace{-3mm}&=&\hspace{-3mm}\left\{
\begin{bmatrix} \frac{2}{3} \\[0.5em] \frac{2}{3} \\[0.5em] 0 \\ 0\end{bmatrix},
\begin{bmatrix} \frac{2}{3} \\[0.5em] \frac{4}{3} \\[0.5em] 0 \\ 0\end{bmatrix},
\begin{bmatrix} \frac{4}{3} \\[0.5em] \frac{2}{3} \\[0.5em] 0 \\ 0\end{bmatrix},
\begin{bmatrix} \frac{4}{3} \\[0.5em] \frac{4}{3} \\[0.5em] 0 \\ 0\end{bmatrix},
\begin{bmatrix} \frac{2}{3} \\[0.5em] 0 \\ \frac{2}{3} \\[0.5em] 0\end{bmatrix},
\begin{bmatrix} \frac{2}{3} \\[0.5em] 0 \\ \frac{4}{3} \\[0.5em] 0\end{bmatrix},
\begin{bmatrix} \frac{4}{3} \\[0.5em] 0 \\ \frac{2}{3} \\[0.5em] 0\end{bmatrix},
\begin{bmatrix} \frac{4}{3} \\[0.5em] 0 \\ \frac{4}{3} \\[0.5em] 0\end{bmatrix}, \right.
\nonumber \\ & & \left. \hspace{-1mm}
\begin{bmatrix} 0 \\[0.5em] \frac{2}{3}  \\ 0\\ \frac{2}{3}\\[0.5em]\end{bmatrix},
\begin{bmatrix} 0 \\[0.5em] \frac{2}{3}  \\ 0\\ \frac{4}{3}\\[0.5em]\end{bmatrix},
\begin{bmatrix} 0 \\[0.5em] \frac{4}{3}  \\ 0\\ \frac{2}{3}\\[0.5em]\end{bmatrix},
\begin{bmatrix} 0 \\[0.5em] \frac{4}{3}  \\ 0\\ \frac{4}{3}\\[0.5em]\end{bmatrix},
\begin{bmatrix} 0\\ 0 \\[0.5em] \frac{2}{3} \\[0.5em] \frac{2}{3}\end{bmatrix},
\begin{bmatrix} 0\\ 0 \\[0.5em] \frac{2}{3} \\[0.5em] \frac{4}{3}\end{bmatrix},
\begin{bmatrix} 0\\ 0 \\[0.5em] \frac{4}{3} \\[0.5em] \frac{2}{3}\end{bmatrix},
\begin{bmatrix} 0\\ 0 \\[0.5em] \frac{4}{3} \\[0.5em] \frac{4}{3}\end{bmatrix}
\right \}. \nonumber
\end{eqnarray}}

{\em Example 2:} Let $N_t=7$ and $N_a=2$. To achieve a transmission
efficiency of 8 bpcu, we need four intensity levels $I_m=\frac{2I_pm}{5}$, 
$m = 1,2,3,4$. In this case, we need only 16 activation patterns out of 
${7 \choose 2}=21$ possible activation patterns. The choice of these 
activation patterns will determine the performance of the GSM system, 
since choosing a particular activation pattern can alter the minimum 
Euclidean distance between any two GSM signal vectors $\vx_1$ and 
$\vx_2$ for a given $\mh$, which is given by
\begin{equation}
d_{min,\mh} \Define\min_{{\vx _1} ,{\vx _2}\in \gsm } \ \|\mh({\vx _2}-{\vx _1})\|^2 .
\label{dmin}
\end{equation}
Similarly, the average Euclidean distance between any two 
vectors $\vx_1$ and $\vx_2$ for a given $\mh$ is given by
\begin{equation}
d_{avg,\mh}=\frac{1}{{|\gsm| \choose 2}}\sum_{{\vx _1} ,{\vx _2}\in \gsm }\big \|\mh({\vx _2}-{\vx _1})\big \|^2 .
\label{davg}
\end{equation}

\subsubsection*{Optimum placement of LEDs in a square grid}
\label{opt_place}
\begin{figure}
\centering
\includegraphics[height=2.75in]{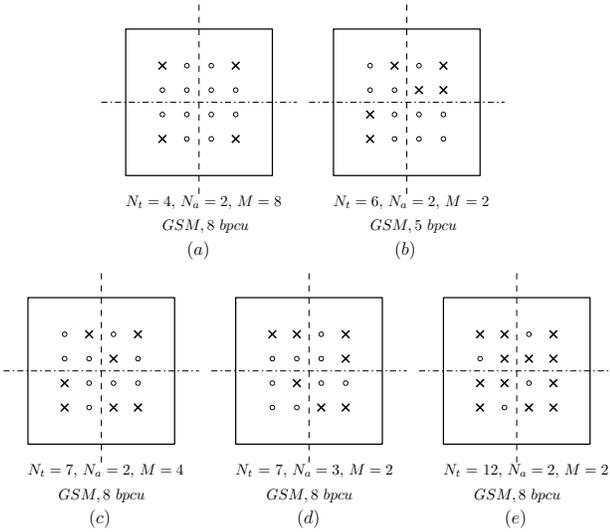}
\caption{Optimum placement of LEDs for GSM in a $4\times 4$ grid. $\times$
indicates the presence of an LED and $\circ$ indicates the absence of LED.}
\label{placements}
\vspace{-2mm}
\end{figure}

Since $d_{min,\mh}$ in (\ref{dmin}) and $d_{avg,\mh}$ in (\ref{davg}) 
influence the link performance, we use them as the metrics based on which 
the optimum placement of LEDs is chosen. Specifically, we choose the 
placement of the LEDs at the transmitter such that the $d_{min,\mh}$ and 
$d_{avg,\mh}$ of the placement are maximized over all possible placements, 
as follows. We first choose the placement(s) for which the $d_{min,\mh}$ 
is maximum. For placement of LEDs in a $p\times q$ grid, we enumerate 
all possible LED placements in the grid and compute the $d_{min,\mh}$ in 
(\ref{dmin}) for all these placements and choose the one with the 
maximum $d_{min,\mh}$. If there are multiple placements for which 
$d_{min,\mh}$ is maximum, we then compute $d_{avg,\mh}$ as per 
(\ref{davg}) for these placements and 
choose the one with the maximum 
$d_{avg,\mh}$. For example, for the system parameters specified in Table 
\ref{tab1} and a required transmission efficiency of 8 bpcu (using 
$N_t=4, N_a=2, M=8$), the best placement of $N_t=4$ LEDs in a $4\times 4$ 
grid that maximizes $d_{min,\mh}$ and $d_{avg,\mh}$ is shown in Fig. 
\ref{placements}(a). Likewise, the best LED placements for systems with  
($N_t=6, N_a=2, M=2$, 5 bpcu), 
($N_t=7, N_a=2, M=4$, 8 bpcu), 
($N_t=7, N_a=3, M=2$, 8 bpcu), and 
($N_t=12, N_a=2, M=2$, 8 bpcu) in
a $4\times 4$ grid are as shown in Figs. \ref{placements}(b),(c),(d),(e), 
respectively. 

\begin{table}
\centering
\begin{tabular}{|l||l|l|} 
\hline 
 	& Length $(X)$ 	& 5m \\ \cline{2-3}
Room   	& Width ($Y$) 	& 5m \\ \cline{2-3}
 	& Height ($Z$) 	& 3.5m	\\ \hline \hline
 	& Height from the floor & 3m \\ \cline{2-3}
 	& Elevation 	& $-90\degree$ \\ \cline{2-3}
Transmitter & Azimuth 	& $0\degree$ \\ \cline{2-3}
 	& $\Phi_{1/2}$ 	& $60\degree$ \\ \cline{2-3}
 	& Mode number, $n$ & 1 \\ \cline{2-3}
 	& $d_{tx}$ 	& 0.6m \\ \hline \hline
 	& Height from the floor & 0.8m\\ \cline{2-3}
 	& Elevation 	& $90\degree$ \\ \cline{2-3}
Receiver & Azimuth 	& $0\degree$ \\ \cline{2-3}
 	& Responsivity, $r$ & 0.75 Ampere/Watt  \\ \cline{2-3}
 	& FOV 		& $85\degree$ \\ \cline{2-3}
 	& $d_{rx}$ 	& 0.1m \\ \hline
\end{tabular}
\vspace{2mm}
\caption{\label{tab1} System parameters in the considered indoor VLC system.}
\end{table}

\section{Performance analysis of GSM in VLC}
\label{sec4}
In this section, we derive an upper bound on the bit error rate (BER) 
of ML detection for GSM in indoor VLC systems. 
The ML detection rule for GSM in the VLC system model described in the
previous section is given by
\begin{equation}
\hat{\vx} =\argmin_{\vx \in \gsm} \ \|\vy-r\mh\vx\|^2.
\label{hatx}
\end{equation}

\subsection{Upper bound on BER}
\label{subsec4a}
Consider the system model in (\ref{sysmodel}). Normalizing the elements
of the noise vector to unit variance, the received vector in (\ref{sysmodel}) 
becomes
\begin{eqnarray}
\vy = \frac{r}{\sigma}\mh\vx +\vn,
\label{rec}
\end{eqnarray} 
and the ML detection rule in (\ref{hatx}) can be rewritten as
\begin{equation}
\hat{\vx} =\argmin_{\vx \in \gsm}\ \big(\frac{r}{\sigma}\|\mh\vx\|^2-2\vy ^T\mh\vx\big).
\label{xhat}
\end{equation}
Assuming that the channel matrix $\mh$ is known at the receiver, the 
pairwise error probability (PEP) -- probability that the receiver decides 
in favor of the signal vector $\vx _2$ when $\vx _1$ was transmitted -- 
can be written as

\vspace{-3mm}
{\small
\begin{eqnarray}
\hspace{-6mm}PEP_{gsm} &\hspace{-3mm}=&\hspace{-2mm}PEP({\vx _1} \rightarrow {\vx_2}|\mh) \nonumber\\
&\hspace{-3mm}=&\hspace{-2mm}P\bigg (\vy ^T\mh({\vx _2}-{\vx _1})> \frac{r}{2\sigma }\big (\|\mh{\vx _2}\|^2 - \|\mh{\vx _1}\|^2\big )\bigg) \nonumber \\
 &\hspace{-3mm}=&\hspace{-2mm}P\bigg (\frac{2\sigma}{r}\vn^T\mh(\vx_2 -\vx_1)>\|\mh(\vx_2 -\vx_1 )\|^2\bigg).
\label{PEP1}
\end{eqnarray}
}

\vspace{-2mm}
\hspace{-5mm}
Define $z\Define\frac{2\sigma}{r}\vn^T\mh(\vx_2 -\vx_1)$. We can see
that $z$ is a Gaussian r.v. with mean $\se (z)=0$ and variance 
$\mbox{Var}(z)=\frac{4\sigma^2}{r^2}\|\mh({\vx _2}-{\vx _1})\|^2$.
Therefore, (\ref{PEP1}) can be written as 
\begin{equation}
PEP_{gsm}=Q\bigg (\frac{r}{2\sigma}\|\mh({\vx _2}-{\vx _1})\|\bigg).
\label{PEPQ}
\end{equation}
Define $\mathcal{A}\Define|\gsm|$. An upper bound on the BER for ML 
detection can be obtained using union bound as 
{\small
\begin{eqnarray}
BER_{gsm} &\hspace{-2mm}\leq&\hspace{-2mm} 
\frac{1}{\mathcal{A}\eta_{gsm}}\sum_{i=1}^{\mathcal{A}}\sum_{j=1,i\neq j}^{\mathcal{A}} d_{H}({\vx _i},{\vx _j}) PEP({\vx _i}\rightarrow {\vx _j}|\mh) \nonumber \\
&\hspace{-25mm}=&\hspace{-13mm}\frac{1}{\mathcal{A}\eta_{gsm}}\sum_{i=1}^{\mathcal{A}}\sum_{j=1,i\neq j}^{\mathcal{A}}\hspace{-1mm} d_{H}({\vx _i},{\vx _j}) Q\bigg (\frac{r}{2\sigma}\|\mh({\vx_j}-{\vx_i})\|\bigg),
\label{BER1}
\end{eqnarray}
}
\hspace{-3.5mm}
where ${d_{H}({\vx _i},{\vx _j})}$ is the Hamming distance between the 
bit mappings corresponding to the signal vectors $\vx _i$ and $\vx _j$. 
Similar BER upper bounds for other MIMO modulation schemes like SMP and SM
have been derived in \cite{vlc5},\cite{vlc6}. We will see in the numerical 
results section next (Sec. \ref{subsec4b}) that the BER upper bound for GSM 
in (\ref{BER1}) is tight at moderate to high SNRs.

\subsection{Numerical results}
\label{subsec4b}
In this section, we present numerical results which illustrate the
tightness of the analytical bound in comparison with the simulated BER
under different system parameter settings. The VLC system parameters 
considered are listed in Table \ref{tab1}. We fix the number of photo 
detectors at the receiver to be $N_r=4$ throughout. 

\subsubsection{Comparison of upper bound and simulated BER}
\label{subsubsec4b1}
\begin{figure}
\centering
\includegraphics[height=2.5in]{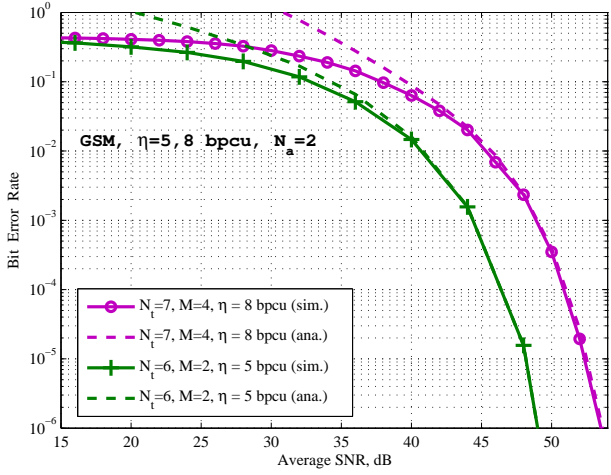}
\caption{Comparison of analytical upper bound and simulated BER
for GSM with ML detection in VLC systems with $i)$ $N_t=6, N_a=2, M=2$,
$\eta=5$ bpcu, and $ii)$ $N_t=7, N_a=2, M=4$, $\eta=8$ bpcu. $N_r=4$.}
\vspace{-2mm}
\label{ubs}
\end{figure}

In Fig. \ref{ubs}, we plot the simulated BER along with the upper bound 
in (\ref{BER1}) for GSM with ML detection in VLC systems with $i)$ $N_t=6$, 
$N_a=2$, $M=2$, $\eta=5$ bpcu, and $ii)$ $N_t=7$, $N_a=2$, $M=4$, 
$\eta=8$ bpcu. The placement of LEDs for these two configurations is 
done over a $4\times4$ grid as depicted in Figs. \ref{placements}(b),(c),
respectively. From the BER plots in Fig. \ref{ubs}, it can be seen that 
the derived upper bound on BER is very tight at moderate to high SNRs,
thus validating the analysis. 

\subsubsection{Comparison of different GSM configurations for fixed $\eta$}
\label{subsubsec4b2}
\begin{figure}
\centering
\includegraphics[height=2.5in]{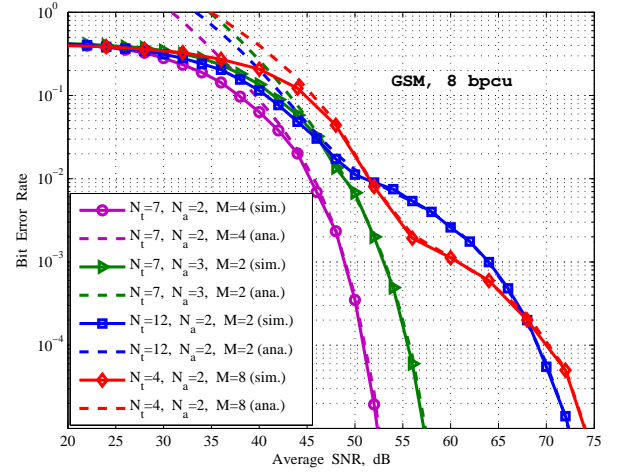}
\caption{Comparison of the BER performance of different configurations of 
GSM with $\eta=8$ bpcu. $N_r=4$.}
\label{diffNt}
\vspace{-2mm}
\end{figure}

Here, we compare the BER performance of four different GSM configurations, 
all having the same transmission efficiency of 8 bpcu. These configurations 
are: System-1 with $N_t=4, N_a=2, M=8$, 
System-2 with $N_t=7, N_a=2, M=4$, 
System-3 with $N_t=7, N_a=3, M=2$, and
System-4 with $N_t=12, N_a=2, M=2$.
The placement of LEDs for these configurations is done over a $4\times4$ 
grid as depicted in Figs. \ref{placements}(a),(c),(d),(e), respectively. 
The simulated BER as well as the analytical upper bound on the BER for
these four configurations are plotted in Fig. \ref{diffNt}. From Fig. 
\ref{diffNt}, it can be seen that System-$2$ configuration achieves the 
best BER performance among all the four systems considered, and System-3
achieves the next best performance. The performance of System-1 and
System-4 are quite poor, particularly at high SNRs. The reason for this
relative performance behavior can be attributed to the fact that System-2 
has the largest $d_{min,\mh}$ and $d_{avg,\mh}$ values, and that Systems-1 
and System-4 have lower $d_{min,\mh}$ and $d_{avg,\mh}$ values, which are 
illustrated 
in Table \ref{tab2}. Also, note that System-2 and System-3 have equal 
number of LEDs. But System-3 sees more interference due to higher number 
of active LEDs, and this results in the poor performance of System-3 
compared to that of System-2, despite System-3 having a lower-order 
modulation alphabet ($M$). In System-4, the average distance between 
the active LEDs is smaller, and, hence, the channel correlation is higher. 
This results in the poor performance of System-4. System-1 has the 
poorest performance because of the modulation order $M$ is the highest
compared to other systems, and it has the smallest $d_{min,\mh}$ and 
$d_{avg,\mh}$ values. The plots in Fig. \ref{diffNt} also show that the
bound is very tight at moderate to high SNRs.

\begin{table}[h]
\centering
\begin{tabular}{|c|c|c|c|}
\hline
System & GSM configuration & $d_{min,\mh}$ & $d_{avg,\mh}$\\\hline\hline
{\scriptsize 1} & {\scriptsize $N_t=4, N_a=2, M=8$} & {\scriptsize $4.623\times10^{-17}$ \hspace{-2mm}} & {\scriptsize $4.520\times10^{-11}$ \hspace{-2mm}} \\ \hline
{\scriptsize 2} & {\scriptsize $N_t=7, N_a=2, M=4$} & {\scriptsize $1.977\times10^{-14}$ \hspace{-2mm}} & {\scriptsize $6.601\times10^{-11}$ \hspace{-2mm}} \\ \hline
{\scriptsize 3} & {\scriptsize $N_t=7, N_a=3, M=2$} & {\scriptsize $1.541\times10^{-14}$ \hspace{-2mm}} & {\scriptsize $6.003\times10^{-11}$ \hspace{-2mm}} \\ \hline
{\scriptsize 4} & {\scriptsize $N_t=12, N_a=2, M=2$} & {\scriptsize $1.346\times10^{-16}$ \hspace{-2mm}} & {\scriptsize $4.842\times10^{-11}$ \hspace{-2mm}} \\ \hline
\end{tabular}
\vspace{2mm}
\caption{\label{tab2} Values of $d_{min,\mh}$, $d_{avg,\mh}$ for different 
GSM configurations with $\eta=8$ bpcu.}
\vspace{-3mm}
\end{table}

\subsubsection{Performance of GSM for varying $d_{tx}$}
\begin{figure}[h]
\centering
\includegraphics[height=2.5in]{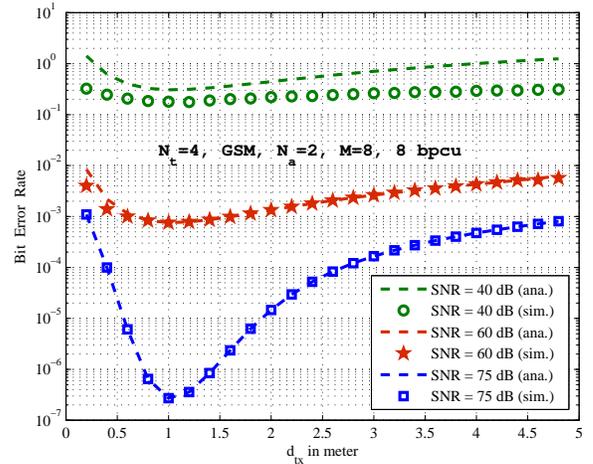}
\caption{BER performance of GSM as a function of $d_{tx}$ in VLC with 
$N_t=4, N_a=2, M=8, \eta=8$ bpcu, $N_r=4$, for different values of 
SNR = 75 dB, 60 dB, 40 dB.}
\label{dtx}
\vspace{-2mm}
\end{figure}

Here, we present the BER performance of GSM in VLC as a function of the 
spacing between the LEDs ($d_{tx}$) by fixing other system 
parameters. 
Figure \ref{dtx} presents the BER performance of GSM as a 
function of $d_{tx}$ in VLC with $N_t=4, N_a=2, M=8, \eta=8$ bpcu, 
for different values of SNR = 75 dB, 60 dB, 40 dB. It can be observed 
from Fig. \ref{dtx} that there is an optimum $d_{tx}$ spacing which 
achieves the best BER performance; below and above this optimum $d_{tx}$
spacing, the BER performance gets worse. The optimum $d_{tx}$ is found to
be 1m in Fig. \ref{dtx}. This optimum spacing can be explained as follows. 
On the one hand, the channel gains get weaker as $d_{tx}$ increases. This 
reduces the signal level received at the receiver, which is a source of 
performance degradation. On the other hand, the channel correlation also 
gets weaker as $d_{tx}$ is increased. This reduced channel correlation is 
a source of performance improvement. These opposing effects of weak channel 
gains and weak channel correlations for increasing $d_{tx}$ leads to an 
optimum spacing. 

\subsubsection{Performance of GSM for varying $\Phi_{1/2}$}
\label{subsubsec4b4}
\begin{figure}
\centering
\includegraphics[height=2.5in]{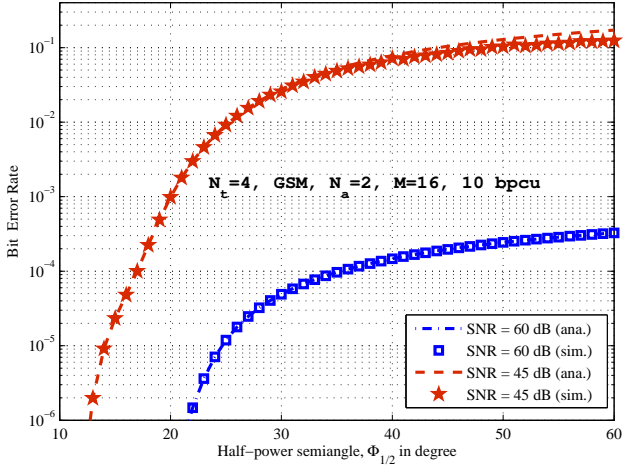}
\caption{BER performance of GSM as a function of $\Phi_{1/2}$ in VLC 
with $N_t=4, N_a=2, M=16, \eta=10$ bpcu, $N_r=4$, $FOV=45\degree$.}
\label{Phi}
\vspace{-2mm}
\end{figure}

Here, we present the effect of varying the half-power semiangle 
($\Phi_{1/2}$) on the BER performance of GSM in VLC. In Fig. \ref{Phi}, 
we present the BER as a function of $\Phi_{1/2}$ in a VLC system
$N_t=4, N_a=2, M=16, \eta=10$ bpcu, and $FOV=45\degree$. BER versus
$\Phi_{1/2}$ plots for SNR = 45 dB, 60 dB are shown. It can 
be observed that the BER performance is good for small half-power 
semiangles, and it degrades as the half-power semiangle is increased. 
This is because, fixing all 
other system parameters as such and decreasing $\Phi_{1/2}$ increases 
the mode number, and hence the channel gain. This increased channel gain
for decreasing $\Phi_{1/2}$ is one reason for improved BER at small
$\Phi_{1/2}$. Another reason is that the channel correlation decreases
as $\Phi_{1/2}$ decreases. This decreased channel correlation also
leads to improved performance at small $\Phi_{1/2}$.

\section{Performance comparison of GSM with other MIMO schemes in VLC}
\label{sec5}
In this section, we compare the performance of GSM with those of other 
MIMO schemes including SMP, SSK, GSSK, and SM, for the same transmission 
efficiency. In all cases, optimum placement of LEDs in a $4\times 4$ grid 
is done based on maximizing $d_{min,\mh}$ and $d_{avg,\mh}$, as described 
in Sec. \ref{opt_place}. 

\begin{figure}[h]
\centering
\includegraphics[height=1.4in]{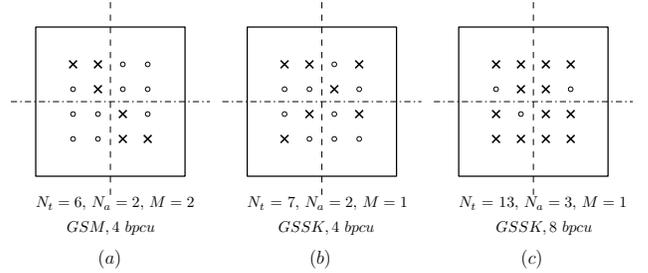}
\caption{Optimum placement of LEDs in a $4\times 4$ grid. $\times$ indicates
the presence of an LED and $\circ$ indicates of absence of LED.}
\label{new_place}
\end{figure}

\begin{figure}[t]
\centering
\includegraphics[height=2.55in]{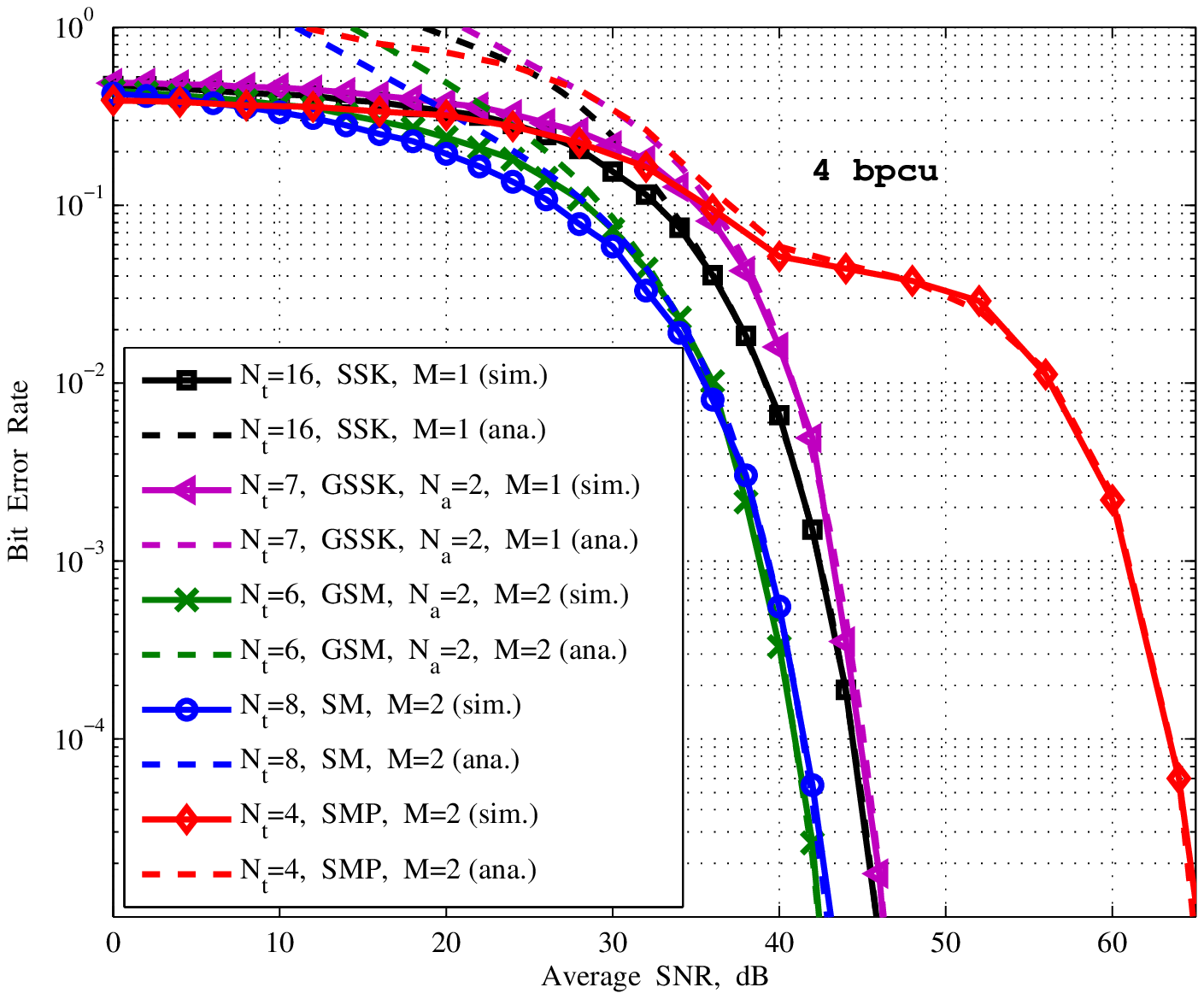}
\caption{Comparison of the BER performance of SMP, SSK, GSSK, SM and GSM
in VLC at $\eta=4$ bpcu. $N_r=4$.}
\label{4bpcu}
\vspace{-2mm}
\end{figure}

In Fig. \ref{4bpcu}, we present the BER performance of SMP, SSK, GSSK, SM, 
and GSM, all having a transmission efficiency of $\eta=4$ bpcu. A GSM 
system with $N_t=6, N_a=2, M=2$ which uses only 4 activation patterns 
chosen out of ${6\choose 2}=15$ activation patterns and gives 4 bpcu is 
considered. The optimum placement of LEDs for this GSM system is as shown 
in Fig. \ref{new_place}(a). The other MIMO schemes with 4 bpcu transmission 
efficiency considered for comparison are:
$i)$ SMP: $N_t=4, N_a=4, M=2$, LEDs placement as in Fig. \ref{placements}(a), 
$ii)$ SSK: $N_t=16, N_a=1, M=1$, LEDs placement as in Fig. \ref{place}(c), 
i.e., one LED on each of the grid point, 
$iii)$ GSSK: $N_t=7, N_a=2, M=1$, LEDs placement as in Fig. \ref{new_place}(b), 
and $iv)$ SM: $N_t=4, N_a=1, M=4$, LEDs placement as in \ref{placements}(a). 
From Fig. \ref{4bpcu}, it can be seen that SM outperforms SMP, which is 
due to spatial interference in SMP. It is also observed that SM performs 
better than SSK and GSSK. This is because SSK has more LEDs and hence the 
$d_{min,\mh}$ and $d_{avg,\mh}$ in SSK are smaller than those in SM. Also, 
in GSSK, 
2 LEDs are activated simultaneously leading to spatial interference, and 
this makes GSSK to perform poorer than SM. Both SSK and GSSK perform better 
than SMP, due to the dominance of spatial interference in SMP. It is further 
observed that GSM performs almost the same as SM, with marginally inferior 
performance at low SNRs (because of the effect of spatial interference in 
GSM) and marginally better performance at high SNRs (because of better 
$d_{min,\mh}$ and $d_{avg,\mh}$ in GSM). 

\begin{figure}
\centering
\includegraphics[height=2.55in]{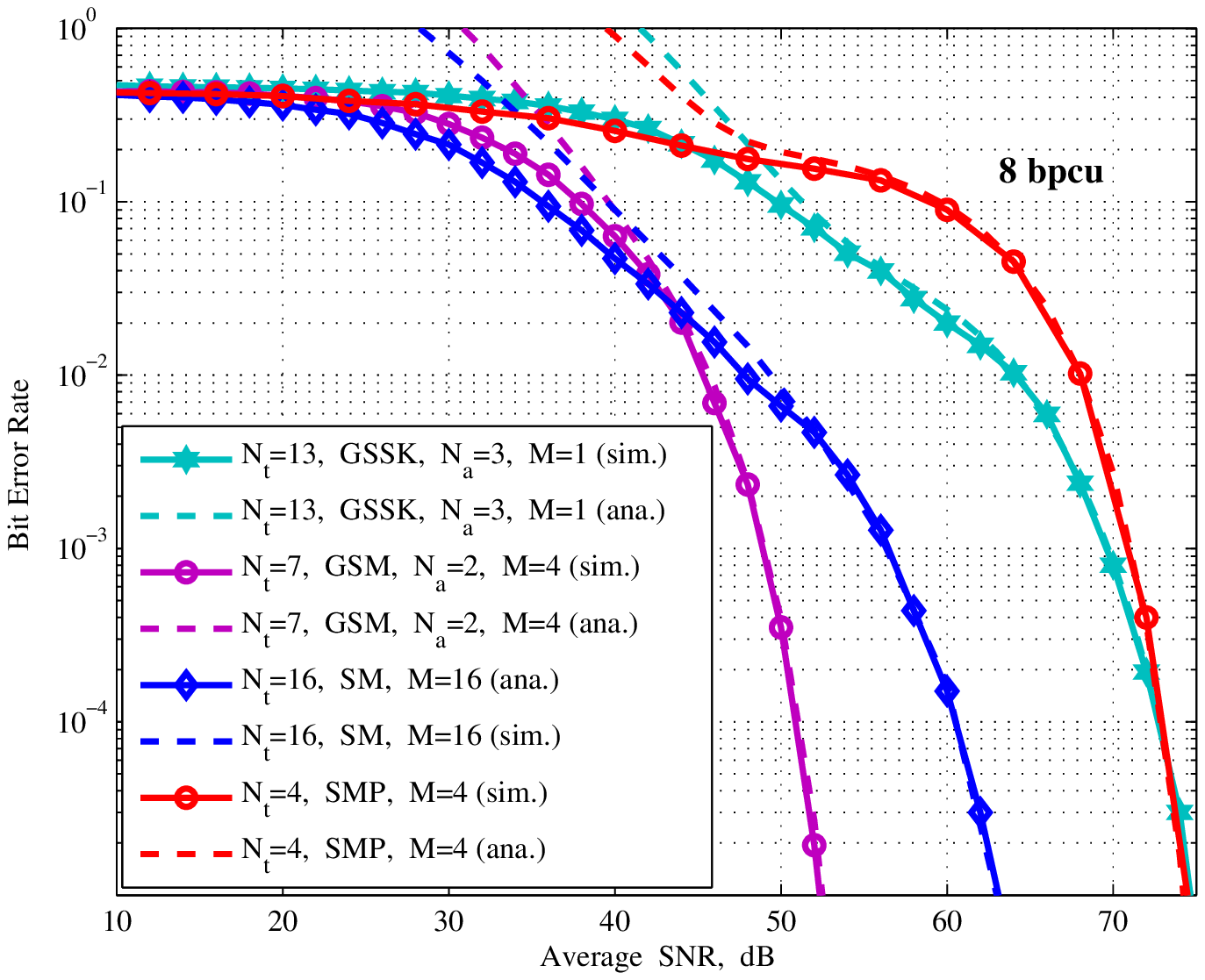}
\caption{Comparison of the BER performance of SMP, GSSK, SM, and GSM in 
VLC at $\eta=8$ bpcu. $N_r=4$.}
\label{8bpcu}
\vspace{-2mm}
\end{figure}

The performance advantage of GSM over SM at high SNRs is substantial at 
8 bpcu transmission efficiency (about 10 dB advantage at $10^{-5}$ BER), 
which is illustrated in Fig. \ref{8bpcu}. Figure \ref{8bpcu} compares
the performance of the following systems, all having 8 bpcu efficiency:
$i)$ SMP: {\small $N_t=4, N_a=4, M=4$}, LEDs placement as in 
\ref{placements}(a), 
$ii)$ GSSK: {\small $N_t=13, N_a=3, M=1$}, LEDs placement as in 
\ref{new_place}(c),
$iii)$ SM: {\small $N_t=16, N_a=1, M=16$}, LEDs placement as in 
\ref{place}(c), and 
$iv)$ GSM: {\small $N_t=7, N_a=2, M=4$}, LEDs placement as in 
\ref{placements}(c). 
From Fig. \ref{8bpcu}, it is observed that GSM achieves the best 
performance among the considered schemes at moderate to high SNRs
(better by about 10 dB compared to SM, and by about 25 dB compared to
GSSK and SMP at $10^{-5}$ BER). The reason for this is as explained 
in the performance comparison in Fig. \ref{4bpcu}.
 
\begin{figure}[h]
\centering
\includegraphics[height=2.5in]{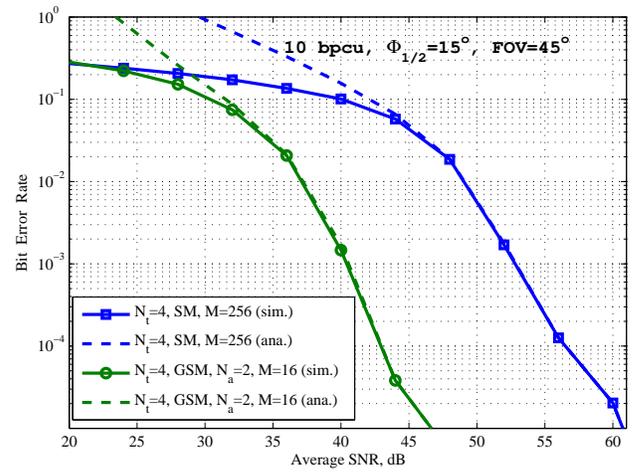}
\caption{Comparison of the BER performance of SM and GSM in VLC at
$\eta=10$ bpcu, $\Phi_{1/2}=15\degree$, $FOV=45\degree$, $N_r=4$.}
\label{10bpcu}
\vspace{-2mm}
\end{figure}

In Fig. \ref{10bpcu}, we compare the BER performance of SM and GSM in VLC, 
both having the same $\eta=10$ bpcu, $\Phi_{1/2}=15\degree$, and 
$FOV=45\degree$. The SM and GSM system parameters are: $i)$ SM: 
{\small $N_t=4, N_a=1, M=256$}, and $ii)$ GSM: 
{\small $N_t=4, N_a=2, M=16$}. The placement
of LEDs in both cases is as in Fig. \ref{placements}(a). It is observed 
that GSM significantly outperforms SM (by 
about 25 dB at $10^{-5}$ BER). This performance advantage of 
GSM over SM can be attributed to the following reasons. The channel matrix 
becomes less correlated for $\Phi_{1/2}=15\degree$, which results in less 
spatial interference in GSM. Despite the presence of multiple active LEDs 
($N_a=2$) and hence spatial interference in GSM, to achieve 10 bpcu 
transmission efficiency, GSM requires a much smaller-sized modulation 
alphabet $(M=16)$ compared to that required in SM $(M=256)$. The better 
power efficiency in a smaller-sized modulation alphabet compared to a 
larger-sized alphabet dominates compared to the degrading effect of 
spatial interference due to $N_a=2$, making GSM to outperform SM. 

\section{Conclusions}
\label{sec6}
We investigated the performance of GSM, an attractive MIMO transmission
scheme, in the context of indoor wireless VLC. More than one among the
available LEDs are activated simultaneously in a channel use, and the 
indices of the active LEDs also conveyed information bits in addition 
to the information bits conveyed by the intensity modulation alphabet.   
To our knowledge, such a study of GSM in VLC has not been reported before. 
We derived an analytical upper bound on the BER of GSM with ML 
detection in VLC. The derived bound was shown to be very tight at 
moderate to high SNRs. The channel gains and channel correlations 
influenced the GSM performance such that the best BER is achieved at 
an optimum LED spacing. Also, the GSM performance in VLC improved
as the half-power semi-angle of the LEDs is decreased.
We compared the BER performance of GSM with those of other MIMO 
schemes including SMP, SSK, GSSK and SM. Analysis and simulation
results revealed favorable performance for GSM compared to other 
MIMO schemes.

\end{document}